\documentclass[sigconf]{acmart}
\pdfoutput=1 
\usepackage{booktabs} 

\usepackage{graphicx}
\usepackage{amsmath}
\usepackage{amsfonts}
\usepackage{amssymb}
\usepackage{xcolor}
\usepackage{enumitem}
\usepackage{url}
\usepackage{listings}
\usepackage{multirow}
\usepackage{colortbl}
\usepackage{hyperref}
\hypersetup{
    colorlinks=true,
    linkcolor=blue,
    filecolor=red,
    urlcolor=magenta,
    breaklinks=true,
}
\usepackage{breakurl}

\copyrightyear{2018} 
\acmYear{2018} 
\setcopyright{acmlicensed}
\acmConference[IWOCL '18]{International Workshop on OpenCL (IWOCL '18)}{May 14--16, 2018}{Oxford, United Kingdom}
\acmBooktitle{IWOCL '18: International Workshop on OpenCL (IWOCL '18), May 14--16, 2018, Oxford, United Kingdom}
\acmPrice{15.00}
\acmDOI{10.1145/3204919.3204924}
\acmISBN{978-1-4503-6439-3/18/05}

\newcommand{\titletext}{CLBlast: A Tuned OpenCL BLAS Library}

\definecolor{commentcolour}{HTML}{888888}
\definecolor{linecolor}{HTML}{888888}
\begin{document}

\title{\titletext}

\author{Cedric Nugteren}
\affiliation{%
  \institution{TomTom}
  \city{Amsterdam}
  \state{The Netherlands}
}
\email{mail@cedricnugteren.nl}

\begin{abstract}

This work introduces CLBlast, an open-source BLAS library providing optimized OpenCL routines to accelerate dense linear algebra for a wide variety of devices. It is targeted at machine learning and HPC applications and thus provides a fast matrix-multiplication routine (GEMM) to accelerate the core of many applications (e.g. deep learning, iterative solvers, astrophysics, computational fluid dynamics, quantum chemistry). CLBlast has five main advantages over other OpenCL BLAS libraries: 1) it is optimized for and tested on a large variety of OpenCL devices including less commonly used devices such as embedded and low-power GPUs, 2) it can be explicitly tuned for specific problem-sizes on specific hardware platforms, 3) it can perform operations in half-precision floating-point FP16 saving bandwidth, time and energy, 4) it has an optional CUDA back-end, 5) and it can combine multiple operations in a single batched routine, accelerating smaller problems significantly.
This paper describes the library and demonstrates the advantages of CLBlast experimentally for different use-cases on a wide variety of OpenCL hardware.

\end{abstract}

\ccsdesc[300]{General and reference~Performance}
\ccsdesc[300]{Theory of computation~Parallel computing models}
\ccsdesc[300]{Computing methodologies~Massively parallel algorithms}
\ccsdesc[300]{Computing methodologies~Parallel programming languages}

\keywords{OpenCL, BLAS, GEMM, Auto-Tuning, GPU, FP16, CUDA, Batched GEMM, HPC, Deep Learning}

\maketitle


\section{Introduction}

Efficient and fast software has become more important than ever as transistor scaling benefits are diminishing~\cite{Borkar2011}, affecting all types of platforms: from embedded devices to desktops and supercomputers. Most of such high performance software is built up around basic building blocks, of which the `Basic Linear Algebra Subroutines' (BLAS) library is one of the most widely used. BLAS is the main dense linear algebra library, providing among others GEMV (generalized matrix-vector multiplication) and GEMM (generalized matrix-multiplication). These routines are nowadays even more important due to their widespread use in deep learning: the most common and compute intensive layers in neural networks are the convolution layers (which can be expressed as the GEMM routine) and the fully-connected layers (either GEMM or GEMV)~\cite{Chetlur2014, Vasudevan2017, Warden2015}. Apart from its new use for deep learning, BLAS remains a pillar for many HPC application areas such as quantum chemistry and fluid dynamics, and for other domains such as machine learning in general, computer vision, and data analytics.

Now, more than 30 years after the introduction of the original Netlib BLAS API, many highly optimized implementations are available for all kinds of purposes and platforms: ATLAS, BLIS, GotoBLAS, OpenBLAS, MKL and so on. However, for graphics processing units (GPUs) and other parallel processors there are fewer alternatives. The most well-known GPU BLAS implementation is NVIDIA's cuBLAS. However, since it is written in CUDA, cuBLAS will not work on non-NVIDIA hardware. Furthermore, it is closed-source. The main alternative is the open-source clBLAS library, written in OpenCL and thus supporting many platforms. However, it is originally designed for AMD GPUs and does not perform well or sometimes does not work at all on other devices which support OpenCL, such as GPUs from NVIDIA and Intel, embedded devices (e.g. Mali, Adreno), FPGAs and CPUs. Moreover, features relevant for deep learning such as half-precision and batched operations are missing in clBLAS.

This paper presents CLBlast, a BLAS library written in OpenCL targeting a wide variety of devices including GPUs. It is open-source, it is written in C++11 and OpenCL, it is well tested on different platforms, and it implements a superset of the BLAS routines. This paper introduces CLBlast and subsequently discusses its five main advantages:

\begin{enumerate}

\item All kernels in CLBlast are highly parameterized and are device-agnostic. That way, they can be auto-tuned for a given OpenCL device through integration of the CLTune auto-tuner~\cite{Nugteren2015}. This results in performance portability, which is demonstrated in this paper by showing matching or superior performance compared to clBLAS on NVIDIA, AMD, Intel and ARM devices.

\item Thanks to integration of an auto-tuner, users can also tune CLBlast for specific problem-sizes. For example, in deep learning, matrices will have a particular shape depending on the configuration of a neural network layer. Using the auto-tuner, performance of CLBlast can be maximized for a specific problem. We demonstrate the benefits of problem-specific tuning experimentally.

\item In contrast to existing OpenCL BLAS libraries, CLBlast also implements half-precision routines using the 16-bit floating-point format (FP16). This reduces storage and bandwidth requirements by a factor two, but also allows for much faster and more energy efficient computations (e.g. around 2x faster GEMM on a Skylake GT2 GPU or Mali GPU).

\item The library internally abstracts the OpenCL API behind the new high-level CLCudaAPI. With this API, porting CLBlast host-code to CUDA is trivial, requiring only a simple header change. CLCudaAPI also provides an OpenCL-to-CUDA kernel header making kernel porting easy as well. Thanks to this, CLBlast has a CUDA back-end as well, making it the first fully-featured CUDA BLAS library which is open-source.

\item CLBlast provides a special interface for batching BLAS routines. As we show in this work, this can yield up to an order of magnitude better performance especially when processing many small vectors or matrices. This is of special interest for deep learning, as multiple smaller operations are typically batched.

\end{enumerate}

\section{Related Work}

From a technical perspective, AMD's clBLAS is the most closely related work: it is also an OpenCL BLAS open-source library. However, it does not have CLBlast's performance portability, problem-specific tuning, FP16 support and batched routines. Furthermore, from a technical perspective it lacks proper testing on less-common devices, it has no C++ interface, it requires a newer version of OpenCL (1.2 instead of 1.1), and its OpenCL kernels are partly generated as strings and thus not easily readable or editable. The original authors are no longer developing clBLAS, but they have re-focused on rocBLAS\footnote{rocBLAS: \url{http://github.com/ROCmSoftwarePlatform/rocBLAS}}: a work-in-progress BLAS library written in HIP that currently only supports a small subset of all BLAS routines and data-types.

NVIDIA's cuBLAS is also related, but it is closed source and CUDA-only and thus does not run on non-NVIDIA hardware. Those two libraries (cuBLAS and clBLAS) are also combined together with additional auto-tuned kernels in the ISAAC project\footnote{ISAAC project: \url{http://github.com/ptillet/isaac}}. However, ISAAC is not a full BLAS library yet, supporting only a few routines so far. However, it does support input-size aware auto-tuning based on a learned tuning parameters model~\cite{Tillet2017}.

Other related OpenCL libraries are clMAGMA, ArrayFire, and ViennaCL, but they focus on higher-level routines such as LAPACK rather than BLAS. From a non-OpenCL perspective, ATLAS is the most relevant work, as it also includes a device-specific auto-tuner.

From a scientific perspective, several works have previously published auto-tuning and optimization approaches for dense matrix-matrix multiplications~\cite{Weber2012,Lai2013,Li2009,Matsumoto2012,Matsumoto2014, Tillet2017}. In fact, the GEMM kernel in CLBlast is based on and evolved from the work by Matsumoto et al.~\cite{Matsumoto2014}.
There are also several recent publications on batched GEMM operations and more in general on optimizing GEMM for small matrices~\cite{Abdelfattah2016,Dong2016,Masliah2016}. Related to the batched operations in CLBlast is also a comparison article for possible standard interfaces~\cite{Relton2016}.

\section{The CLBlast Library}

CLBlast is an APACHE 2.0 licensed open-source\footnote{CLBlast repository: \url{http://github.com/CNugteren/CLBlast}} OpenCL implementation of the BLAS API. The host code is written in C++11 and the kernel code in OpenCL C, compatible with any device supporting the OpenCL 1.1 or newer standards. There are automated build tests on Windows, macOS and Linux systems and there is continuous integration through automated correctness tests on six different devices from three different vendors. Furthermore, there are Python bindings in the form of PyCLBlast. CLBlast has an active community: there are third party Java bindings (JOCLBlast) and Nim bindings (nimCLBlast), 11 contributors, 50+ forks, 100+ resolved issues, it is the standard OpenCL BLAS back-end for ArrayFire, it is being used in experimental OpenCL versions of Caffe\footnote{Caffe with CLBlast: \url{http://github.com/dividiti/ck-caffe}} and Tensorflow\footnote{OpenCL Tensorflow: \url{http://github.com/hughperkins/tensorflow-cl}}~\cite{Perkins2017}, and it is used in PyTorch through libgpuarray\footnote{Libgpuarray: \url{http://deeplearning.net/software/libgpuarray}}.

The CLBlast project was created as a stand-alone project rather than by extending and improving the existing clBLAS project, because the kernels in clBLAS are generated from C++ code, making them very difficult to read, extend and maintain. Furthermore, the lack of development by clBLAS's authors and the use of good coding practices contributed to this decision.

\subsection{Library Design}

Implementing the exact Netlib BLAS API would require internal host-to-device and device-to-host OpenCL transfers in the library. This could be detrimental for performance, especially for $\mathcal{O}(n)$ and $\mathcal{O}(n^2)$ BLAS routines. That is why clBLAS and cuBLAS take pointers to device memory in the API, leaving full control over data transfers to the user. For the same reasons, CLBlast also provides this as the main interface. There is a C, C++ and Java interface available. On top of this, there is also a fully compatible Netlib BLAS interface, but this is not recommended for performance since OpenCL data-transfers are done internally, leaving no control to the user.

BLAS routines are divided into three levels. CLBlast implements all of these routines plus a few extra, see table~\ref{tbl:routines}: 10 level-1 scalar, vector, and vector-vector routines, 23 level-2 matrix-vector routines, 9 level-3 matrix-matrix routines, and 9 extra BLAS-like routines. For each of these routines, CLBlast has (if possible) an implementation in 5 different precisions: half-precision FP16 (e.g. HGEMM), single-precision FP32 (e.g. SGEMM), double-precision FP64 (e.g. DGEMM), complex single-precision 2xFP32 (e.g. CGEMM), and complex double-precision 2xFP64 (e.g. ZGEMM).

\begin{table}[!ht]
  \centering
  \caption{Routines available in CLBlast}
  \begin{tabular}{c|l}
    level & routines \\
    \hline
    1 & AXPY COPY SCAL SWAP \\
      & AMAX ASUM DOT DOTC DOTU NRM2 \\
    \hline
    2 & GBMV GEMV HBMV HEMV HPMV SBMV \\
      & SPMV SYMV TMBV TPMV TRMV TRSV \\
      & GER GERC GERU HER HER2 HPR HPR2 \\
      & SPR SPR2 SYR SYR2 \\
    \hline
    3 & GEMM HEMM HER2K HERK SYMM \\
      & SYR2K SYRK TRMM TRSM  \\
    \hline
    extra & SUM MAX MIN AMIN OMATCOPY IM2COL \\
      & AXPYBATCHED GEMMBATCHED \\
      & GEMMSTRIDEDBATCHED
  \end{tabular}
  \label{tbl:routines}
\end{table}

Although there are 51 routines per precision in CLBlast, there are not that many OpenCL kernels implemented. First of all, the kernels are precision-agnostic. Although C++ templates aren't supported in OpenCL C 1.1, we can still use a type alias and at kernel-compile-time define the type as either half, single, double precision or one of the complex data-types. Second, there are several families of kernels which can be re-used for other routines. The kernels \texttt{axpy}, \texttt{dot}, \texttt{gemv}, \texttt{ger} and \texttt{gemm} span already most of the routines given some support kernels such as copying or padding vectors and matrices. For example, to implement the GBMV routine, CLBlast uses the \texttt{gemv} OpenCL kernel but uses a pre-processor macro to change the loading of the input from a general matrix into a banded matrix. The remainder of the kernel with all performance-critical optimizations can be re-used, avoiding code duplication. This is done for almost all routines, similarly to what is described in~\cite{Matsumoto2014} for level-3 routines.

\subsection{Parameterized Kernels}

All kernel implementations in CLBlast are written in a highly parameterized way to be tunable across devices: they are not written for a specific device nor OpenCL implementation. This is achieved by creating pre-processor constants which can be changed without affecting the correctness of the program. A simplified example of the \texttt{axpy} kernel in figure~\ref{fig:axpy_example} illustrates this: the local work-group size is tunable (WGS), the amount of work-per-thread can vary (WPT), and the vector width is configurable (VW). In contrast to clBLAS, we rely on the target compiler to perform the low-level optimizations such as loop-unrolling and pointer-arithmetic, increasing the readability, portability, and maintainability of the kernels.

\begin{figure}[!h]
\begin{lstlisting}
  #define WGS 64 // The local work-group size
  #define WPT 4  // The amount of work-per-thread
  #define VW 2   // Width of vectors X and Y

  typedef float dtype; // Example data-type
  #if VW == 1
   typedef float dtypeV;
  #elif VW == 2
   typedef float2 dtypeV;
  #endif // and similarly for VW == {4, 8, 16}
  
  __kernel __attribute_(reqd_work_group_size(WGS))
  void Xaxpy(const int n, const dtype alpha,
             const __global dtypeV* restrict xgm,
             __global dtypeV* ygm) {
   #pragma unroll
   for (int w = 0; w < WPT; ++w) {
    int i = w * get_global_size(0) + get_global_id(0);
    ygm[i] = ygm[i] + alpha * xgm[i];
   }
  }
\end{lstlisting}
\caption{Simplified example of a parameterized OpenCL kernel from CLBlast.}
\label{fig:axpy_example}
\end{figure}

Although it is unfeasible to discuss all of CLBlast's kernels and their parameters here, we believe that it is important to briefly illustrate CLBlast's generality with a larger parameterized kernel as well: the \texttt{gemm} kernel. Similar to~\cite{Matsumoto2012}, we make many assumptions on the input arguments, which are handled by pre-processing and post-processing kernels. These assumptions are e.g. matrix sizes are a multiple of the work-group sizes, offsets are zero, and matrix B is transposed. This is a good solution for larger problem sizes since $\mathcal{O}(n^2)$ data movement is typically cheaper than $\mathcal{O}(n^3)$ computation, but the hidden constant starts to play a role for smaller $n$. Therefore, there is also a single-kernel `direct' version available for those cases, but it shares most of the design and parameters as discussed below.

\begin{figure}[!t]
  \centering
  \includegraphics[width=1.08\columnwidth]{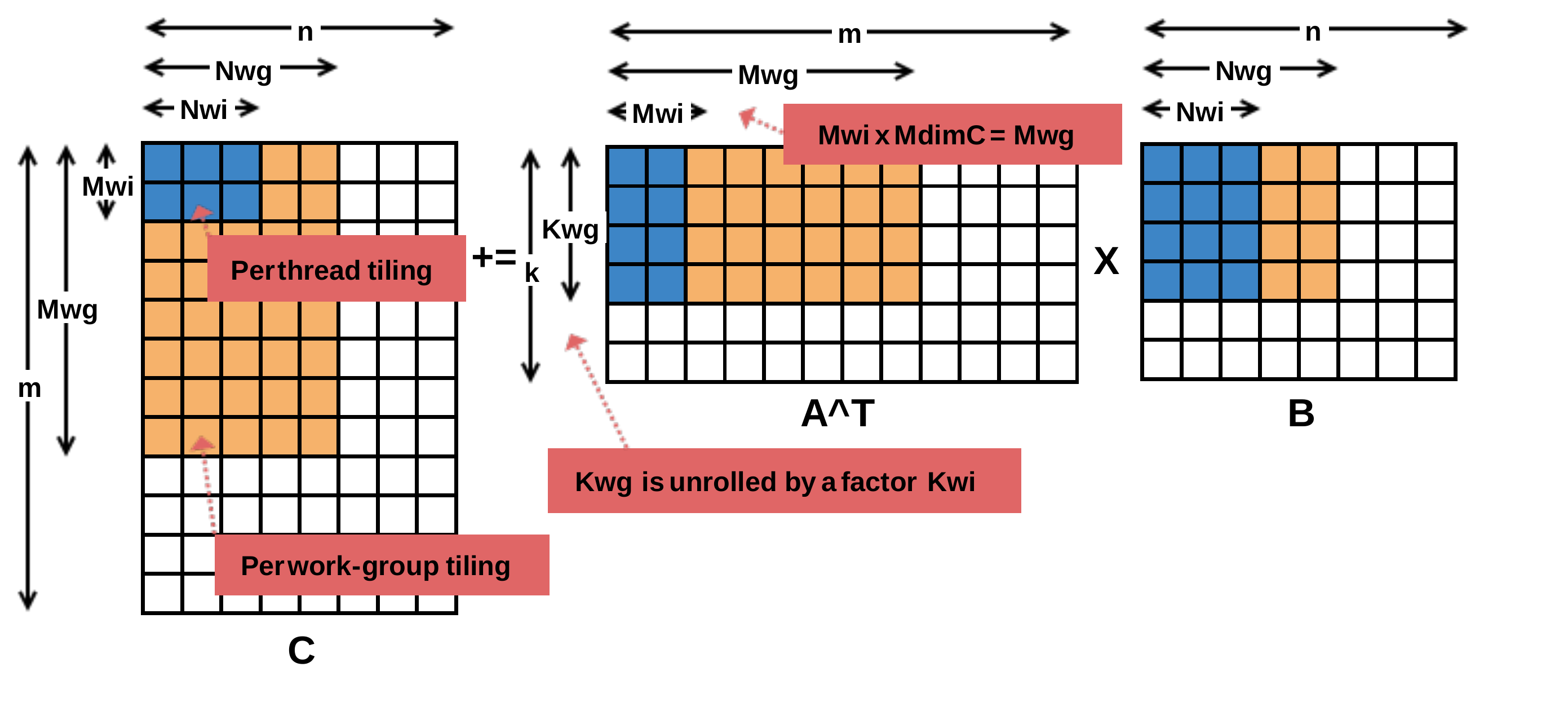}
  \caption{Matrix-multiplication and some of its tuning parameters. The blue area indicates work done by a single thread (`work-item'), the orange area indicates work done per work-group.}
  \label{fig:gemm1}
\end{figure}

The \texttt{gemm} kernel has 14 different parameters, of which 6 are illustrated in figure~\ref{fig:gemm1}. The parameters define among others the work-group sizes in 2 dimensions ($M_{wg}, N_{wg}$), the 2D register tiling configuration ($M_{wi}, N_{wi}$), the vector widths of both input matrices, loop unroll factors ($K_{wi}$), and whether or not and how to use the local memory. For more details we refer to the CLTune paper which discusses an earlier version of the kernel~\cite{Nugteren2015}, and also to the work of Matsumoto et al. which served as inspiration for the design of the kernel~\cite{Matsumoto2012}.

\subsection{Performance Tuning}

The parameterized kernels in CLBlast can be tuned for a specific device and/or specific problem size using an integrated version of the CLTune library\footnote{CLTune: \url{http://github.com/CNugteren/CLTune}}, which is an open-source CUDA and OpenCL auto-tuner written in C++. For details on the CLTune library, we refer to~\cite{Nugteren2015}. CLBlast provides binaries to interface with the auto-tuner for each kernel. By default, these tuners will find optimal kernel parameters for each kernel, for each precision, and for a fixed problem size. Tuning results for previously unseen devices are collected in a central tuning database\footnote{DB: \url{http://github.com/CNugteren/CLBlast-database}} from which CLBlast takes its optimized parameters. The database contains timings of each kernel execution while tuning, not just the optimal. Thus, the database can be used for other purposes as well, such as research on performance modeling or optimal parameter prediction.

To illustrate how the tuners in CLBlast work, consider the \texttt{gemm} kernel discussed earlier. Although each of the parameters has at most only 4 or 5 reasonable values to try, the total search-space explodes quickly and has more than 100.000 possible combinations to explore for the 14 parameters. This is even after filtering out non-valid parameter combinations due to device or software restrictions (e.g. maximum work-group size, local memory size). Depending on the use-case the amount of combinations might be too much to explore. Therefore, CLBlast defines two sets of tuning parameters: one set with the most likely combinations (e.g. 500) and one set with all combinations. The first set is explored exhaustively, while the other set is additionally explored by random sampling in the search-space. The tuner will thus always explore the basic kernel parameter combinations and on top of that an extra user-configurable amount.

All kernels in CLBlast have already been tuned for around 50 different devices thanks to the community, and they can be tuned for any new device. Nevertheless, the library can also perform decently on previously unseen devices: default parameters per device vendor/type and even per device architecture are computed by taking the average best performing parameters for similar devices. For example, kernel parameters for a new unseen AMD GPU will be set to the parameters corresponding to the average best performing case across all existing AMD GPU devices in the database or across all device of that specific architecture if already present. (e.g Tahiti, Vega). Of course, there is no guarantee that performance on the new device will be good or that the parameters are legal at all. However, in case the device is significantly different, the user can always still run the tuners on his/her device to make sure performance is optimized.

The default tuning results are only for a pre-defined problem size to limit the total run-time of the tuners. Nevertheless, users can still tune for their specific problem size, e.g. a small rectangular matrix of size 279 by 32. The CLBlast API furthermore provides an interface to change the library's default parameters at run-time. Thus, the user can also programmatically provide optimized parameters for (multiple) custom problem sizes and/or devices.

\section{Results}

This section contains several experimental results and details on the setup used in this paper. All results and graphs shown are also available on-line\footnote{On-line appendix of this paper with more graphs at:\\ \url{http://cnugteren.github.io/clblast/}}, including results for the same and other devices.

\subsection{Experimental Setup}

This section explains the set-up used in this paper. In this paper we report results of single-precision and half-precision. We test on the OpenCL devices listed in table~\ref{tbl:devices}.

\begin{table}[!ht]
  \small
  \setlength{\tabcolsep}{4pt}
  \centering
  \caption{Overview of the tested OpenCL devices}
  \begin{tabular}{l|l|l|c|}
    vendor and                & archi-    & library     & GFLOPS        \\
    device name               & tecture   & and SDK     & and GB/s      \\
    \hline
    NVIDIA GTX 750Ti          & Maxwell   & CUDA 8.0     & 1305 --  88       \\
    NVIDIA Titan X            & Pascal    & CUDA 9.0     &10974 -- 480       \\
    ARM Mali T628             & Midgard   & Mali r12p0   &   38 --  15       \\
    AMD Radeon HD7970         & Tahiti    & APP 3.0      & 3789 -- 264       \\
    AMD Radeon M370X          & GCN 1     & Apple 2.4.2  & 1024 --  72       \\
    Intel Skylake ULT GT2     & Iris      & Beignet 1.3  &  384 --  30       \\
    Intel Core i5-6200U       & Skylake   & Intel        &   73 --  30       \\
  \end{tabular}
  \label{tbl:devices}
\end{table}

All experimental results presented in the following sections are fully reproducible: they are obtained through the official CLBlast `clients': binaries which compare run-time of CLBlast against other libraries. The clients perform a warm-up run first, followed by 10 repeated timed runs, and output the final graphs as shown in this paper directly. The reported results are based on the average time over those runs. We report GB/s for routines which are typically bandwidth-bound (level-1 and level-2) and GFLOPS for routines which are typically compute-bound (level-3). We test with CLBlast release 1.3.0 (latest as of January 2018) and compare against:
\begin{enumerate}
  \item AMD's clBLAS 2.12 (latest version as of January 2018). After installing clBLAS, we run the included `clBLAS-tune' to fine-tune performance. However, for some devices the tuner did not complete successfully.
  \item NVIDIA's cuBLAS, included as part of the CUDA installation (see table~\ref{tbl:devices} for the version).
\end{enumerate}

\subsection{Performance Across Devices}

\begin{figure*}[!p]
  \centering
  \includegraphics[width=0.68\columnwidth]{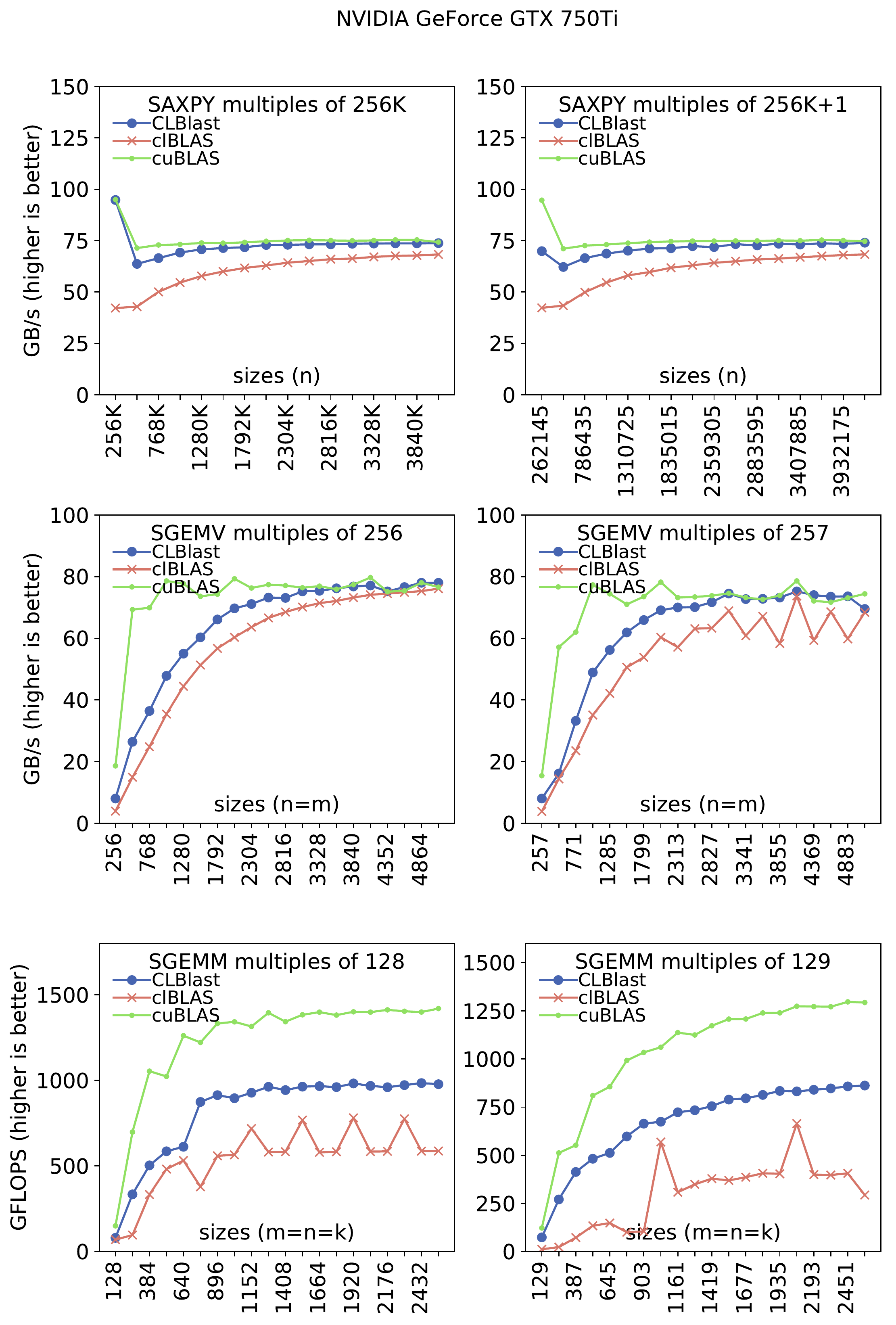}
  {\color{linecolor}\vline}
  \includegraphics[width=0.68\columnwidth]{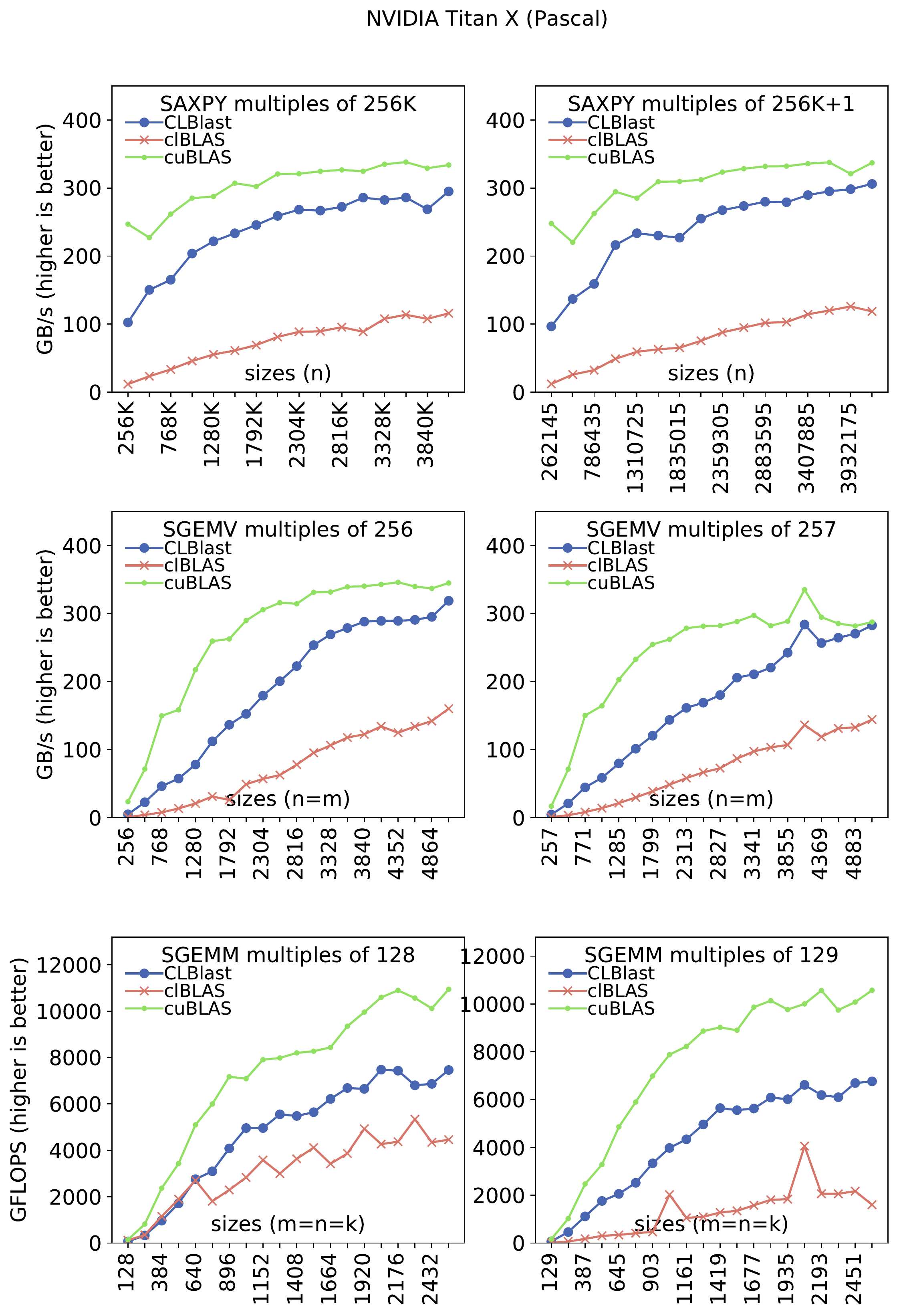}
  {\color{linecolor}\vline}
  \includegraphics[width=0.68\columnwidth]{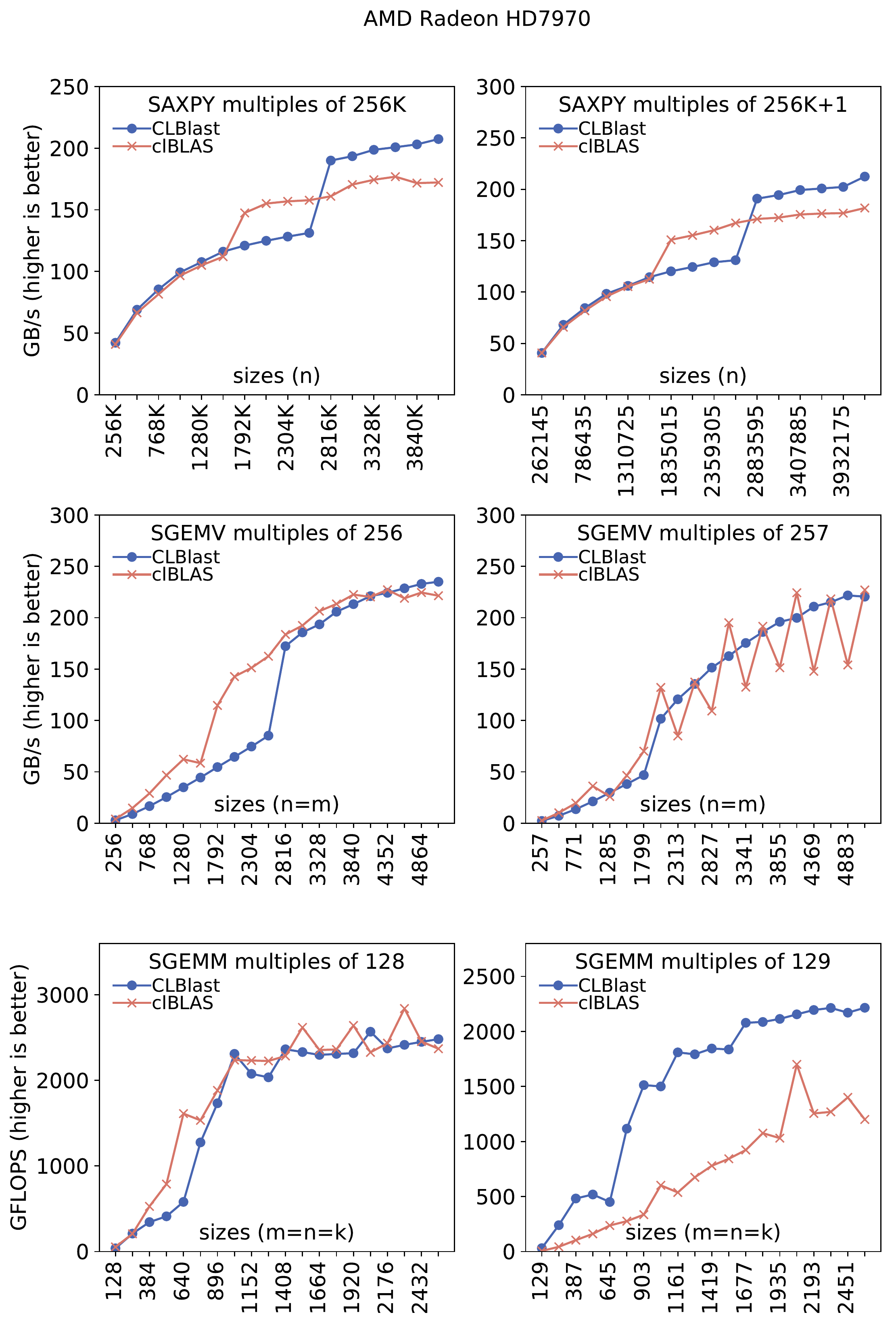}
  \vspace{5pt}
  {\color{linecolor}\hrule}
  \vspace{8pt}
  \includegraphics[width=0.68\columnwidth]{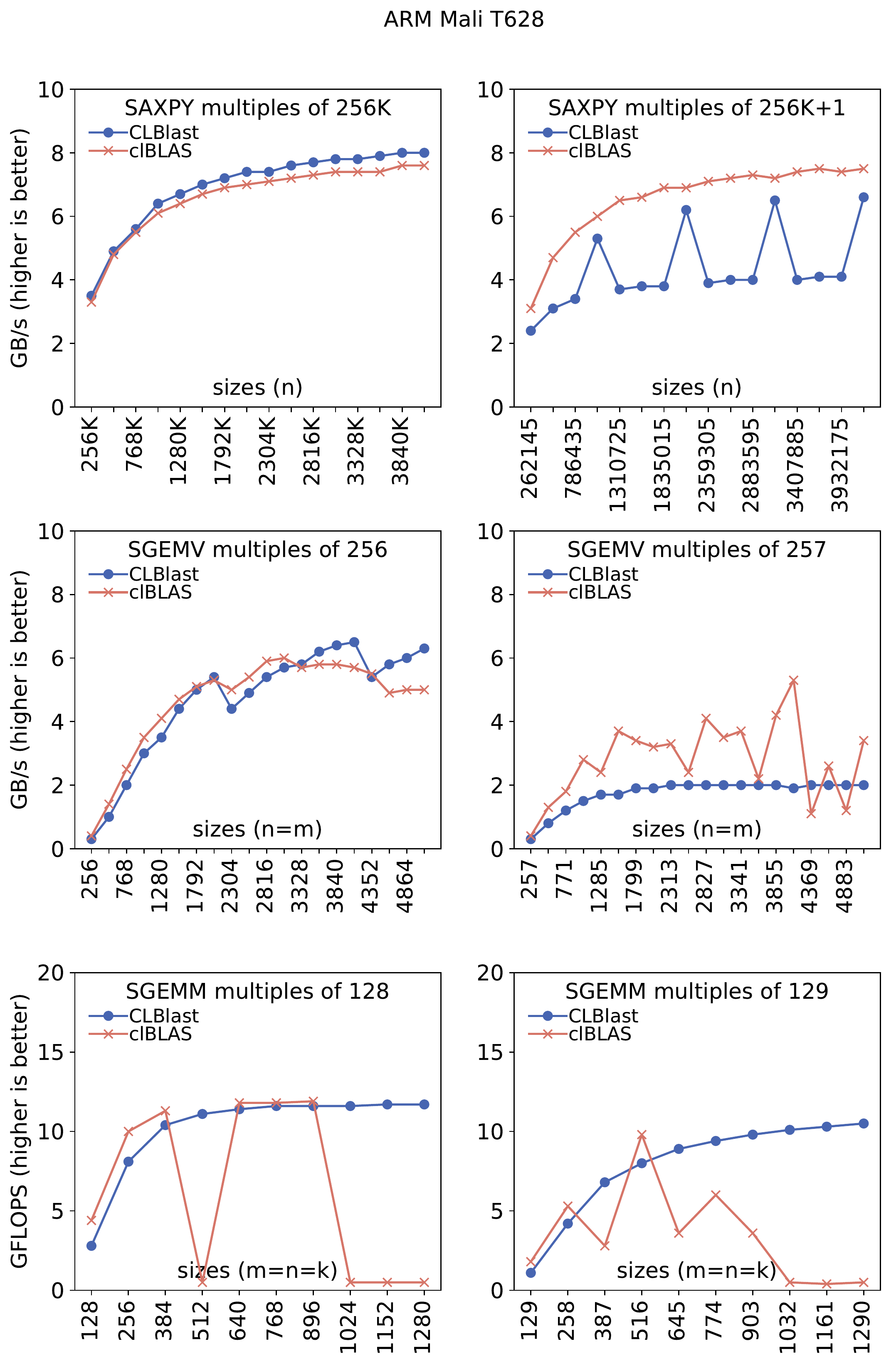}
  {\color{linecolor}\vline}
  \includegraphics[width=0.68\columnwidth]{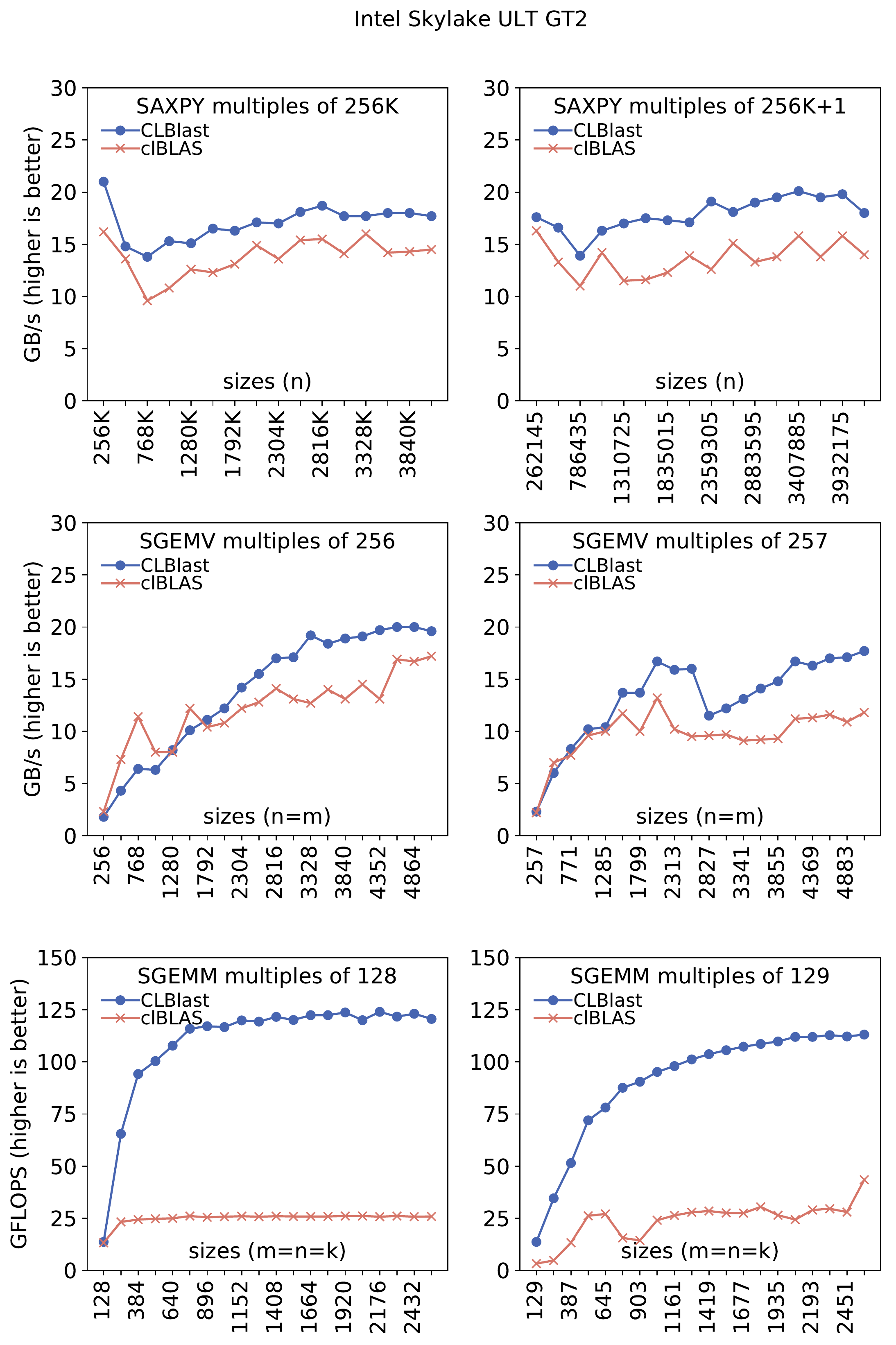}
  {\color{linecolor}\vline}
  \includegraphics[width=0.68\columnwidth]{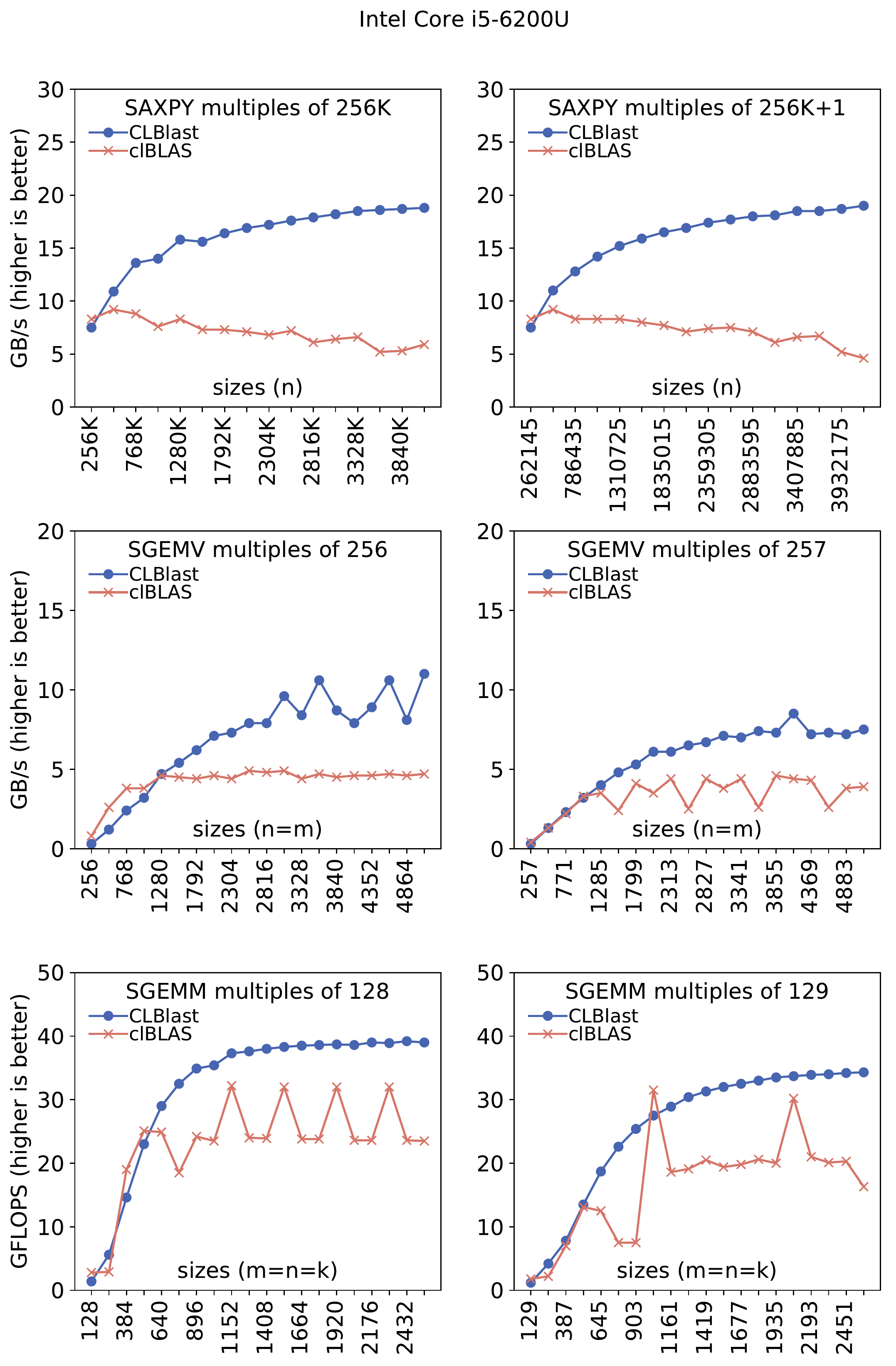}
  \vspace{5pt}
  \caption{Performance across 6 different devices. For each device we show three routines (one per row): SAXPY (measured in GB/s), SGEMV (measured in GB/s) and SGEMM (measured in GFLOPS). For each routine we show two graphs (one per column): left for multiples of a power-of-2, right for multiples of an odd number. Blue circles denote results of CLBlast, red crosses denote results of clBLAS, and green dots denote results of NVIDIA's cuBLAS where available. Best viewed on a computer screen or in the on-line appendix of this paper at~\url{http://cnugteren.github.io/clblast/}.}
  \label{fig:summary}
\end{figure*}

CLBlast is performance-portable due to its tuning capabilities and the generic OpenCL kernels. However, the level of performance achieved is of course still limited by the design and flexibility of the implemented kernels. The design of CLBlast has mainly focused on the \texttt{gemm} kernel which is used for almost all level-3 BLAS operations.

\begin{figure*}[!t]
  \centering
  \includegraphics[width=1.03\columnwidth]{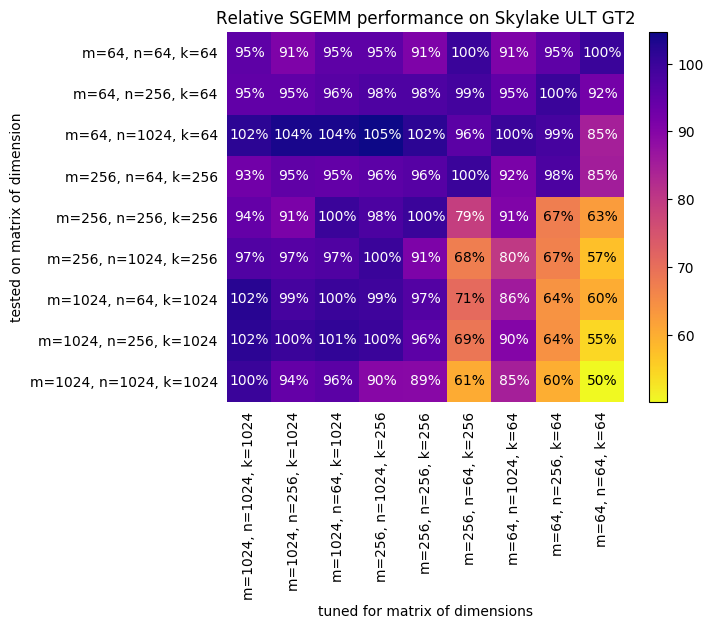}
  {\color{linecolor}\vline}
  \includegraphics[width=1.03\columnwidth]{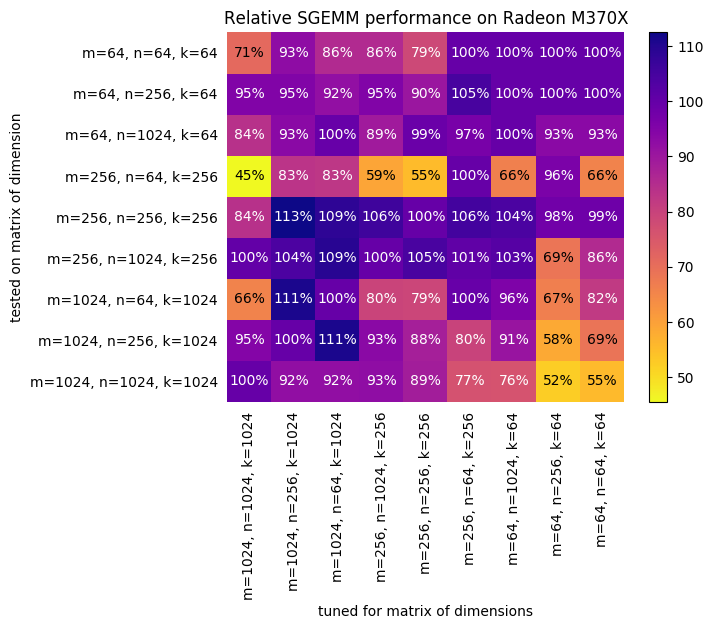}
  \caption{CLBlast performance of SGEMM for different matrix sizes (rows) relative to a matrix-size specific tuned version (diagonal) on two devices. Higher percentages represent faster kernels.}
  \label{fig:heatmaps}
\end{figure*}

In a first set of experiments in figure~\ref{fig:summary} we demonstrate performance-portability across devices. We test on 6 very different devices (see table~\ref{tbl:devices}) for 3 different types of routines: AXPY, GEMV, and GEMM. These routines are chosen for being the most representative for each BLAS level and actually include kernels covering almost all routines. We show results for both multiples of a power-of-2 and for a variety of irregular vector and matrix sizes: multiples of some odd number. For the full results including more complete experiments we refer to the on-line appendix$^9$.
We draw the following conclusions from figure~\ref{fig:summary}:
\begin{itemize}
  \item The AXPY results are for 4 of the 6 devices roughly on-par with clBLAS and cuBLAS. This is to be expected, as it is a simple bandwidth-bound operation. CLBlast tunes the work-group size, the amount of work-per-thread, and the vector width, but the first two don't matter too much for most devices as long as they don't take extreme values. For the CPU experiment this matters more, in which CLBlast is much closer to the peak memory bandwidth. On the Titan X, CLBlast is far ahead of clBLAS and almost matches cuBLAS.
  \item The GEMV results also show on-par or slightly better performance compared to clBLAS and cuBLAS for most devices. More elaborate tests in the on-line appendix$^9$ show that CLBlast's GEMV routine is still sub-optimal: clBLAS achieves better performance for certain cases and devices. Nevertheless, in certain cases CLBlast is much faster, such as on the CPU.
  \item The GEMM results for CLBlast are much more stable across input sizes compared to clBLAS. This is due to the `indirect' kernel design with the additional kernels to transform data in the expected format. The main benefit of this is manifested for irregular sizes.
  \item The GEMM results show significantly better overall results for CLBlast compared to clBLAS. This is especially the case for irregular sizes and for certain devices: Skylake ULT GT2 and the NVIDIA GPUs. However, we note that clBLAS on the Skylake GPU couldn't be tuned due to repeated crashes. Nevertheless, all other devices for which clBLAS worked as expected also show results in favor of CLBlast. This is even true on the tested AMD GPU (for which clBLAS was created), as shown in the multiples-of-129 experiment (bottom-right graphs).
  \item NVIDIA's cuBLAS is still superior over both OpenCL libraries. Because cuBLAS is closed source, we can only formulate hypotheses. First, cuBLAS might be tuned at assembly/PTX level for specific hardware, whereas CLBlast relies on the compiler performing low-level optimizations. Second, specific instructions such as \texttt{\_\_ldg} for L1 data caching are available from CUDA, but not from OpenCL C. A more in-depth analysis of CUDA vs OpenCL for GEMM can be found on-line\footnote{GEMM tutorial: \url{http://www.cedricnugteren.nl/tutorial/}}.
\end{itemize}

In general, performance improvement over clBLAS can be attributed to several factors. For example, clBLAS has more low-level optimizations hard-coded (e.g. pointer arithmetic), leaving less room for the device's compiler to optimize. Furthermore, the tuning space per kernel is limited compared to CLBlast. Examples for GEMM include the lack of a tunable loop unroll factor, no support for strided loading, and absence of the possibility to cache in local memory. Also for GEMM in particular, clBLAS does not pre-transpose the B matrix, resulting in non-subsequent memory accesses for certain tuning parameter configurations.

\subsection{Problem-specific Tuning}

Thanks to the included auto-tuners and the user community, CLBlast is already tuned for a wide variety of devices. However, tuning is currently only done for a default set of input arguments. For example, the GEMM routine is per-default tuned for squared matrices of dimensions $m=1024$, $n=1024$, $k=1024$. To get the maximum performance out of CLBlast, it is possible to tune for specific routine arguments. It is even possible to tune for multiple cases: CLBlast provides an API to change the tuning parameters at run-time. Problem-specific tuning can be beneficial for example for deep learning applications, in which matrices have a specific size based on the neural network's layout.

To illustrate the benefits of problem-specific tuning we tuned the \texttt{gemm} kernel for 9 different matrix sizes. We then benchmarked the SGEMM routine for each of the 9 different sets of tuning parameters on the same 9 problems. The results are shown as heat-maps in figure~\ref{fig:heatmaps}. Shown are the relative performances compared to the diagonal, i.e. the case for which the parameters were tuned. There are a few cases with performance slightly higher than the diagonal (i.e. higher than 100\%), which can happen because the tuner explores a random sub-set of the search space and might have been more fortunate in a particular case. Overall we see quite some potential benefit for problem-specific tuning, even up to a factor 2 for the Radeon M370X GPU: performance drops to $\pm50\%$ if not properly tuned. For the Skylake ULT GT2 GPU we see less benefit: tuning for the larger matrix dimensions seems to generalize towards smaller matrices (top left corner has high values), whereas the opposite is not true (bottom right corner has low values). In conclusion, benefits from problem-specific tuning vary per use-case and per device. In general, it seems definitely worth exploring this problem-specific performance potential.

\begin{figure*}[!t]
  \centering
  \includegraphics[width=1.03\columnwidth]{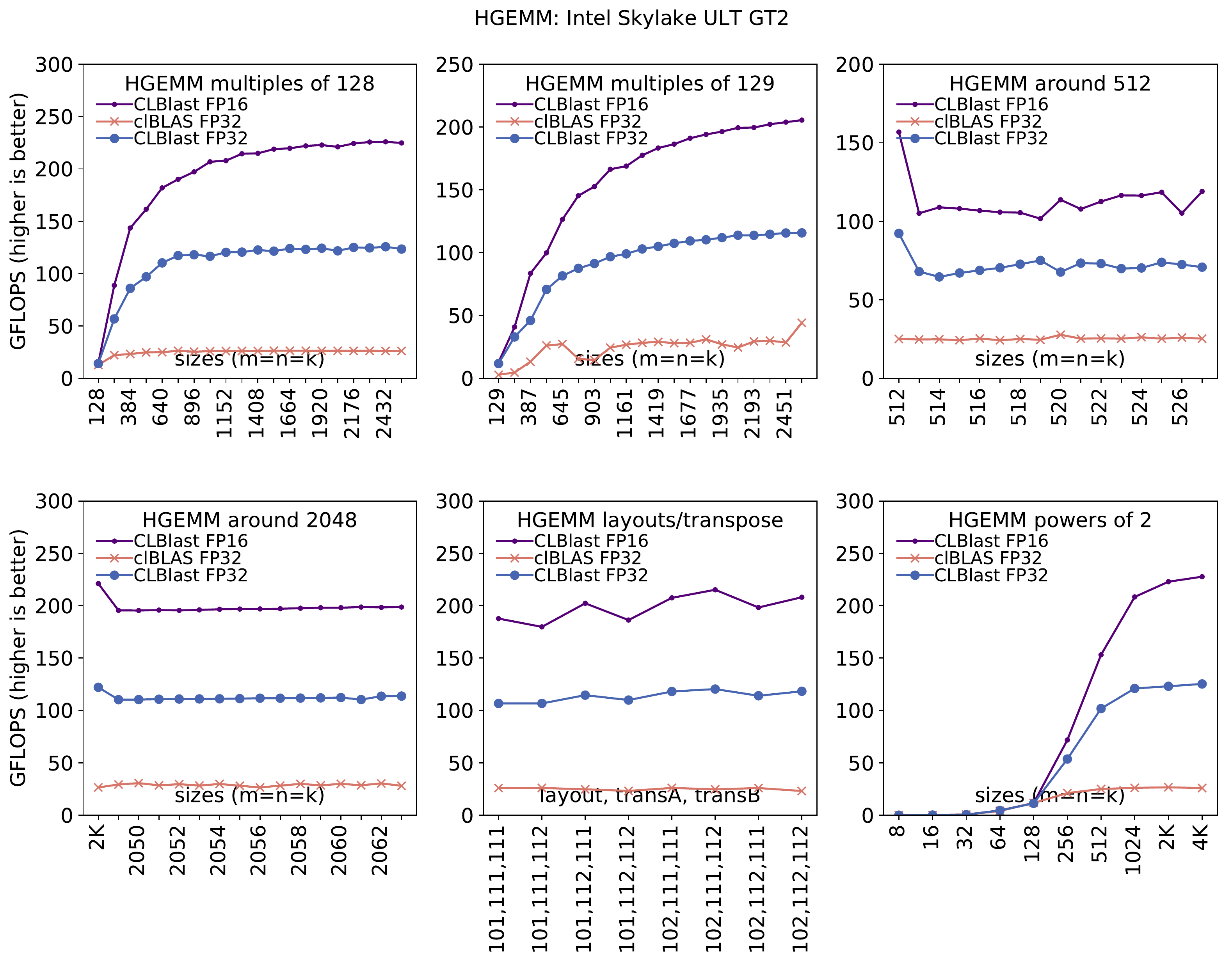}
  {\color{linecolor}\vline}
  \includegraphics[width=1.03\columnwidth]{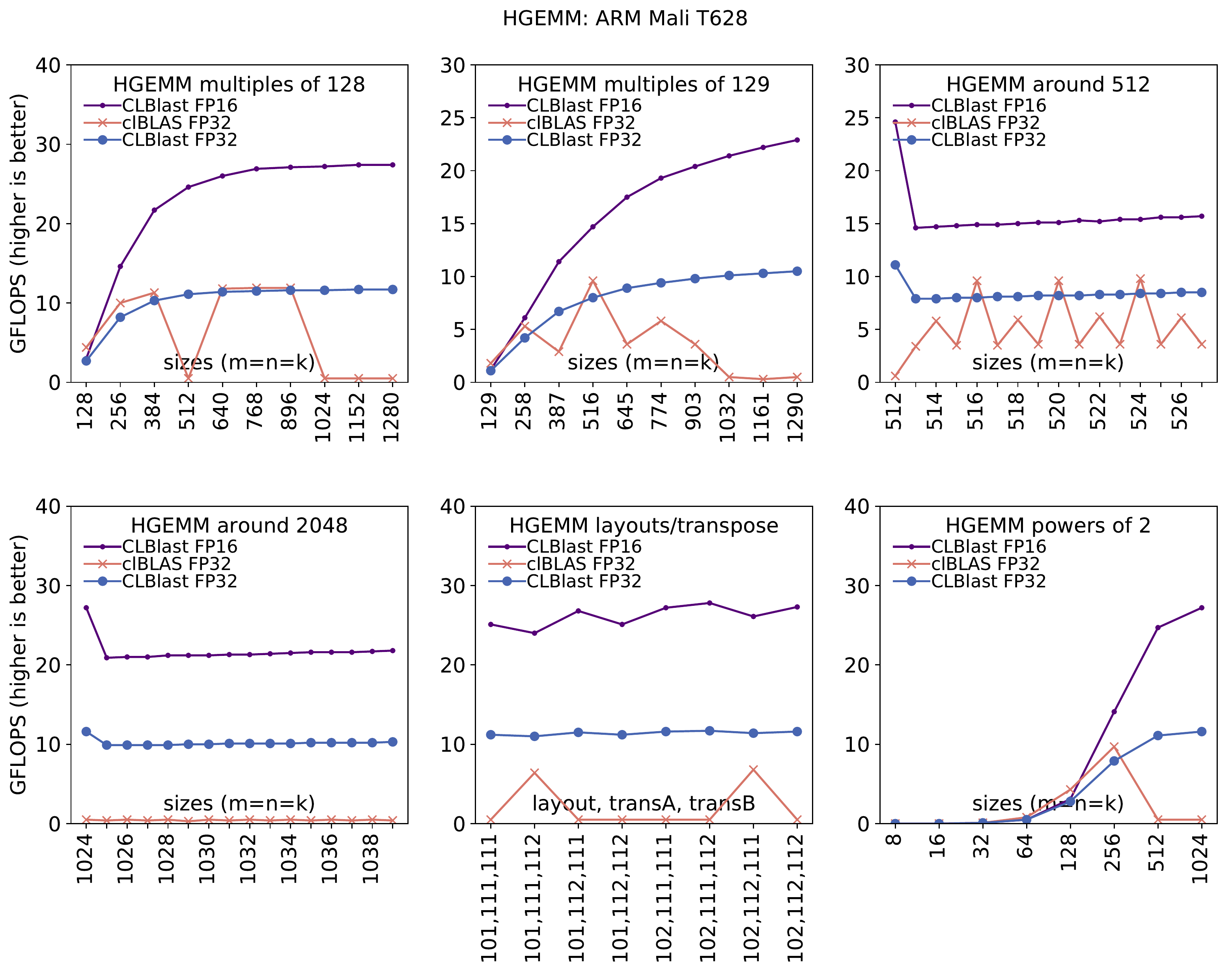}
  \caption{Performance of half-precision matrix-multiplication (HGEMM) on two low-power devices with native FP16 support. Shown are results for CLBlast's FP16 mode (purple dots), CLBlast's FP32 mode (blue circles) and clBLAS's FP32 mode (red crosses).}
  \label{fig:hgemm}
\end{figure*}

\subsection{Half-precision Floating-point}

We also demonstrate the benefit of half-precision floating-point (FP16) in CLBlast. This mode trades-off precision for potential memory savings, faster computation, and less energy consumption. Traditionally used in computer graphics and image processing applications, FP16 has seen renewed interest with the recent successes of deep-learning~\cite{Micikevicius2017}. Devices with native FP16 at 2x FP32 speed can be found in the embedded and low-power domain (e.g. Intel Skylake GPUs and ARM Mali GPUs) and the very high-end (e.g. NVIDIA Tesla P100 and V100). Recent AMD GPUs (Polaris and Vega) also support FP16, but for memory and energy savings only: they run computations at FP32 speed.

CLBlast supports all routines in half-precision mode, while other libraries are lagging behind the recent hardware advances and software requirements. For example, clBLAS has no FP16 support at all. NVIDIA's cuBLAS only supports the GEMM routine in half-precision. Intel does have a single special-purpose OpenCL FP16 kernel integrated in the ISAAC library\footnote{Intel HGEMM \url{http://github.com/ptillet/isaac/pull/20}}, but there is no BLAS library or interface.

We have tested CLBlast's half-precision mode on the two devices with FP16 support from table~\ref{tbl:devices}. Since there is no FP16 support in clBLAS, we test against FP32 versions of CLBlast and clBLAS to show the advantage of FP16. We show results for half-precision GEMM (HGEMM) in figure~\ref{fig:hgemm}, chosen since it is FLOPS-bound rather than bandwidth-bound. From the figure, we can see that for both the Skylake and Mali low-power GPUs achieve around 2x speed-up over CLBlast's FP32 mode, benefiting fully from the hardware's capabilities. The fact that this is sometimes beyond the theoretical 2x is explained by the randomness in the tuning parameter exploration. It is worth noting that on the Skylake ULT GT2 we now achieve over 200 GFLOPS, which is quite an achievement given that is on a laptop system-on-chip sharing 15W with a dual-core Core i5-6200U.

\subsection{CUDA Back-end for CLBlast}

The CLBlast library was originally written with OpenCL as a back-end (hence its name). However, all OpenCL library calls in the host-code were abstracted for convenience through CLCudaAPI, a small header-only C++11 project\footnote{CLCudaAPI: \url{http://github.com/CNugteren/CLCudaAPI}}. This API abstracts away low-level details of OpenCL C calls behind modern C++ classes. Examples are a `Buffer' and a `Kernel' class, with methods such as `Buffer.Write()' or `Kernel.Launch()'. This has several advantages, such as being C++ rather than C, having a higher level interface, benefiting from type templates, and included error checking. However, the main advantage lies in the fact that there is also a CUDA version of CLCudaAPI with an identical interface. Thus, any application written using CLCudaAPI can switch its host-code from OpenCL to CUDA or vice-versa by simply changing a single header `\#include'. This is made possible with the CUDA driver API and NVRTC (since CUDA 7.5) for run-time kernel compilation.

CLBlast includes a compile-time option to easily switch between OpenCL and CUDA host-code without much trouble. However, the kernel code is still written in OpenCL C dialect. Luckily, CLCudaAPI also has a solution for kernel code: an OpenCL-to-CUDA header of around 50 lines of code translates OpenCL kernel code to CUDA using several pre-processor defines and inline functions. This header does not cover the full OpenCL kernel language specification, but it is good enough for the purposes of CLBlast.

\begin{figure*}[!t]
  \centering
  \includegraphics[width=1.03\columnwidth]{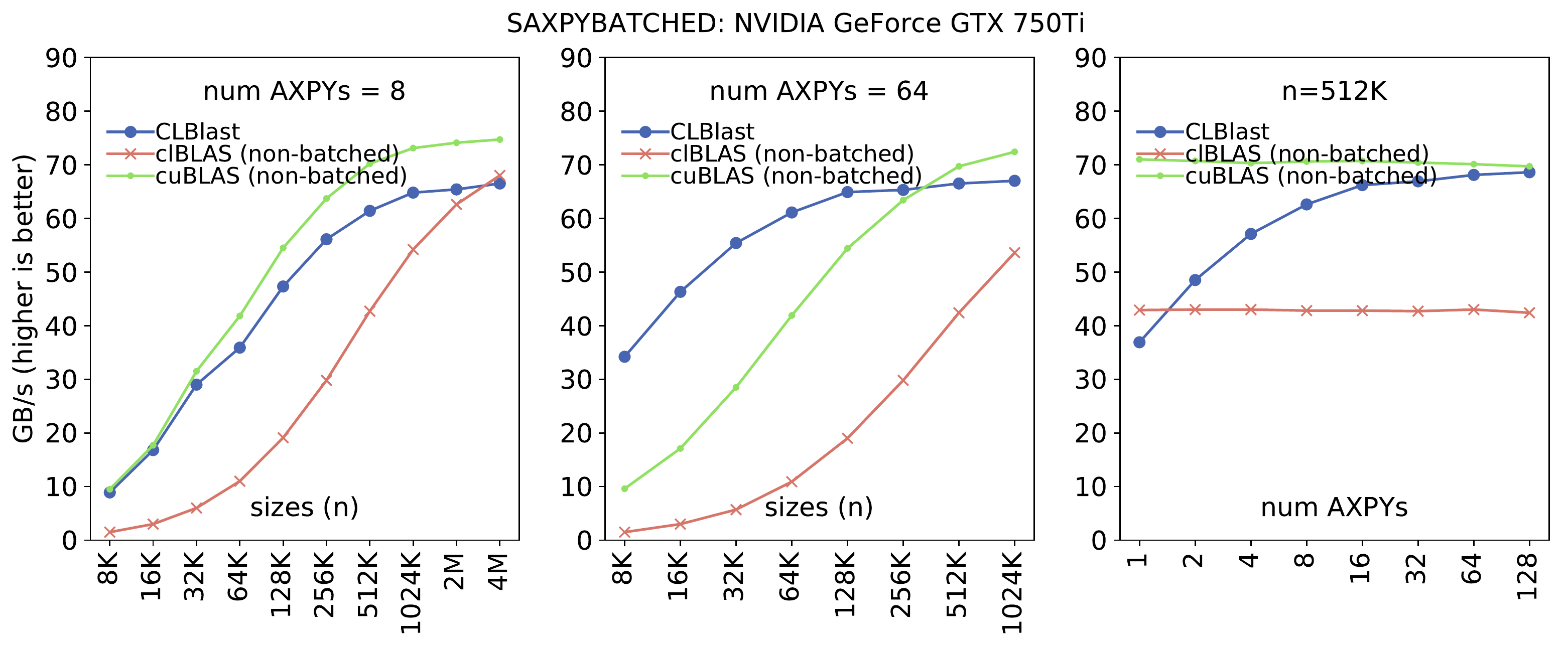}
  {\color{linecolor}\vline}
  \includegraphics[width=1.03\columnwidth]{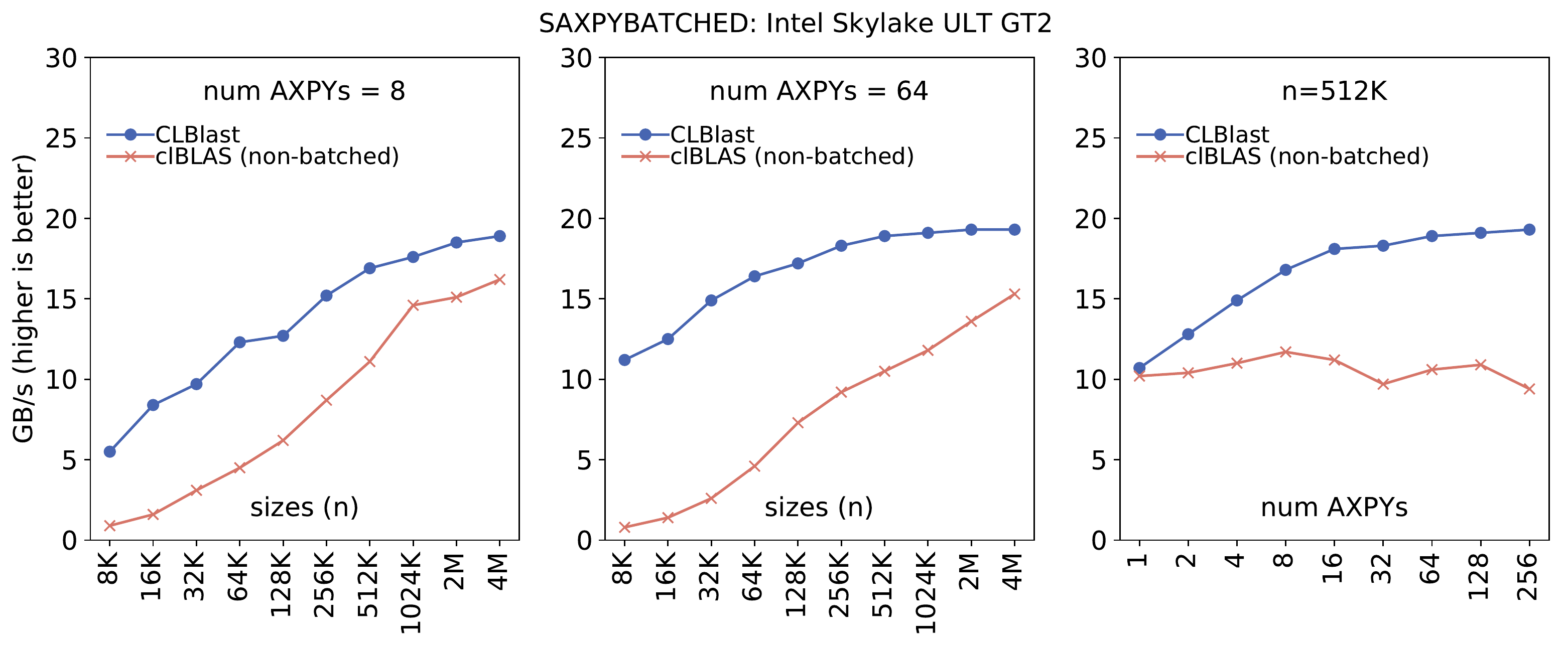}
  \vskip6pt{\color{linecolor}\hrule}\vskip6pt
  \includegraphics[width=1.03\columnwidth]{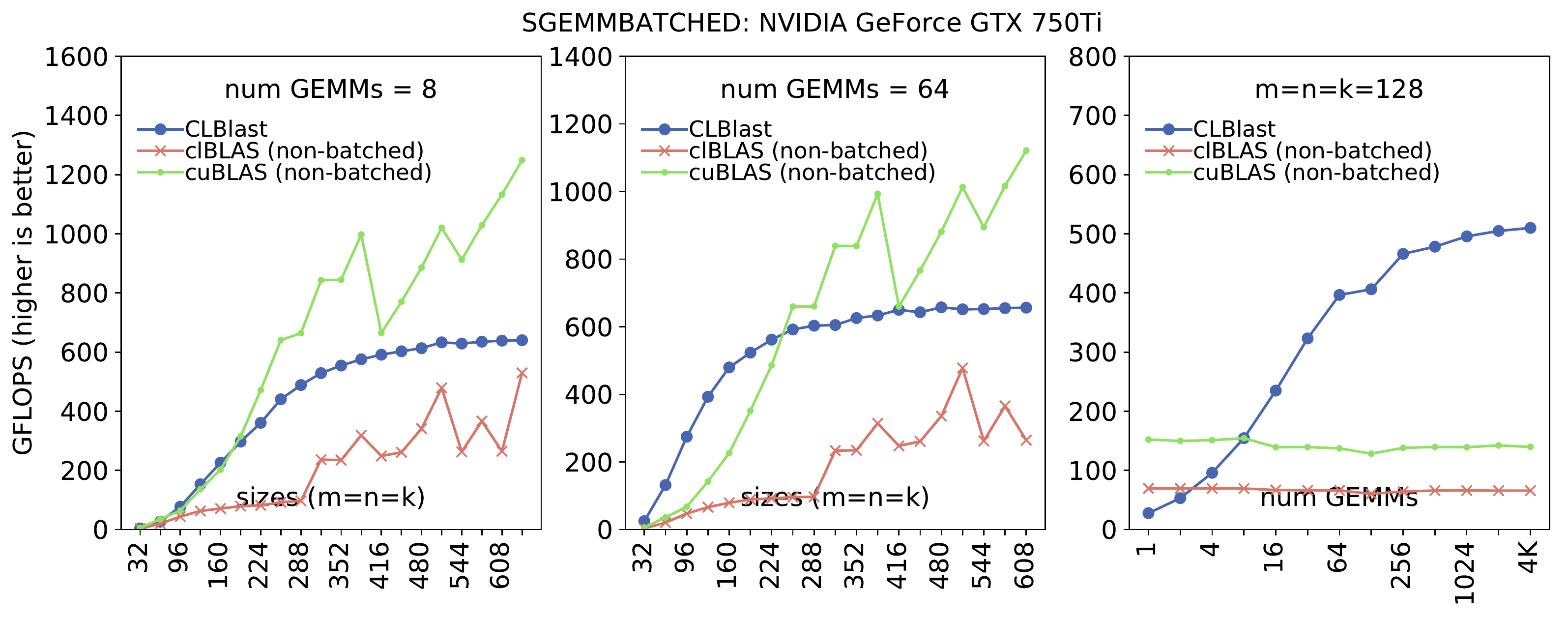}
  {\color{linecolor}\vline}
  \includegraphics[width=1.03\columnwidth]{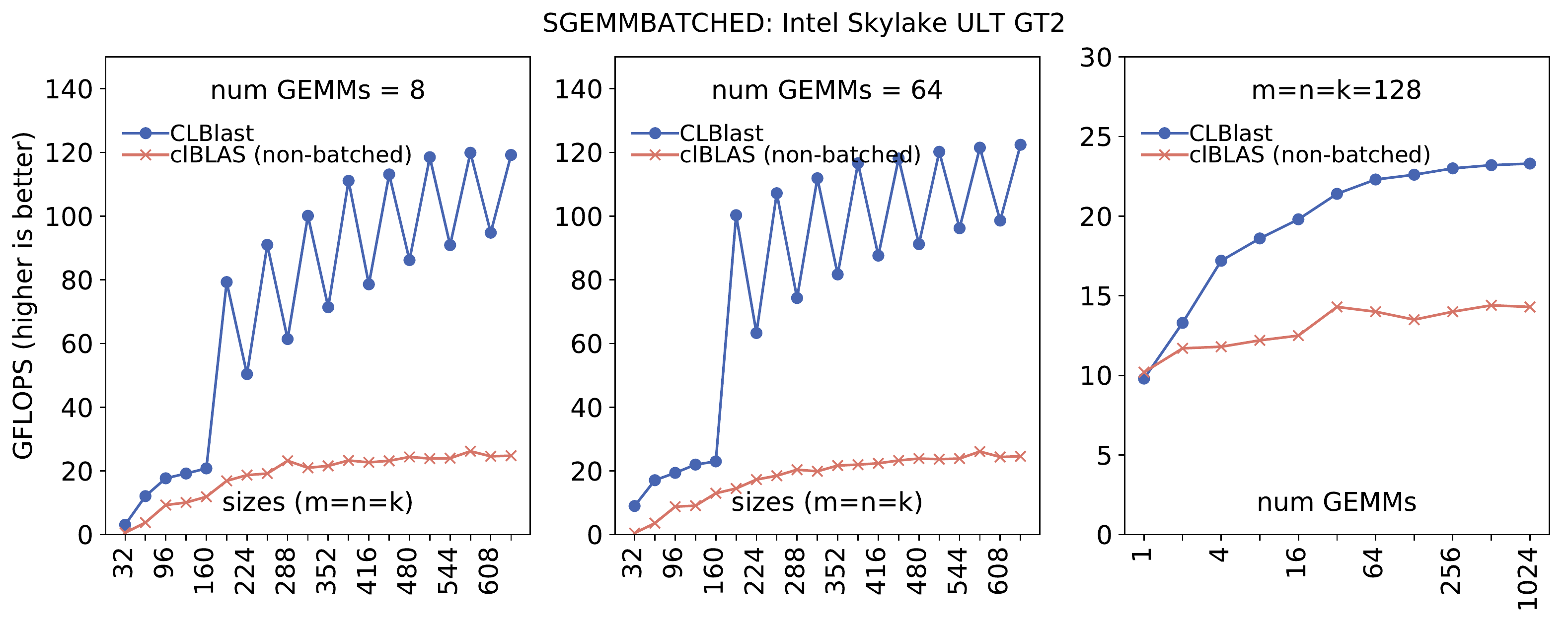}
  \vskip6pt{\color{linecolor}\hrule}\vskip6pt
  \includegraphics[width=1.03\columnwidth]{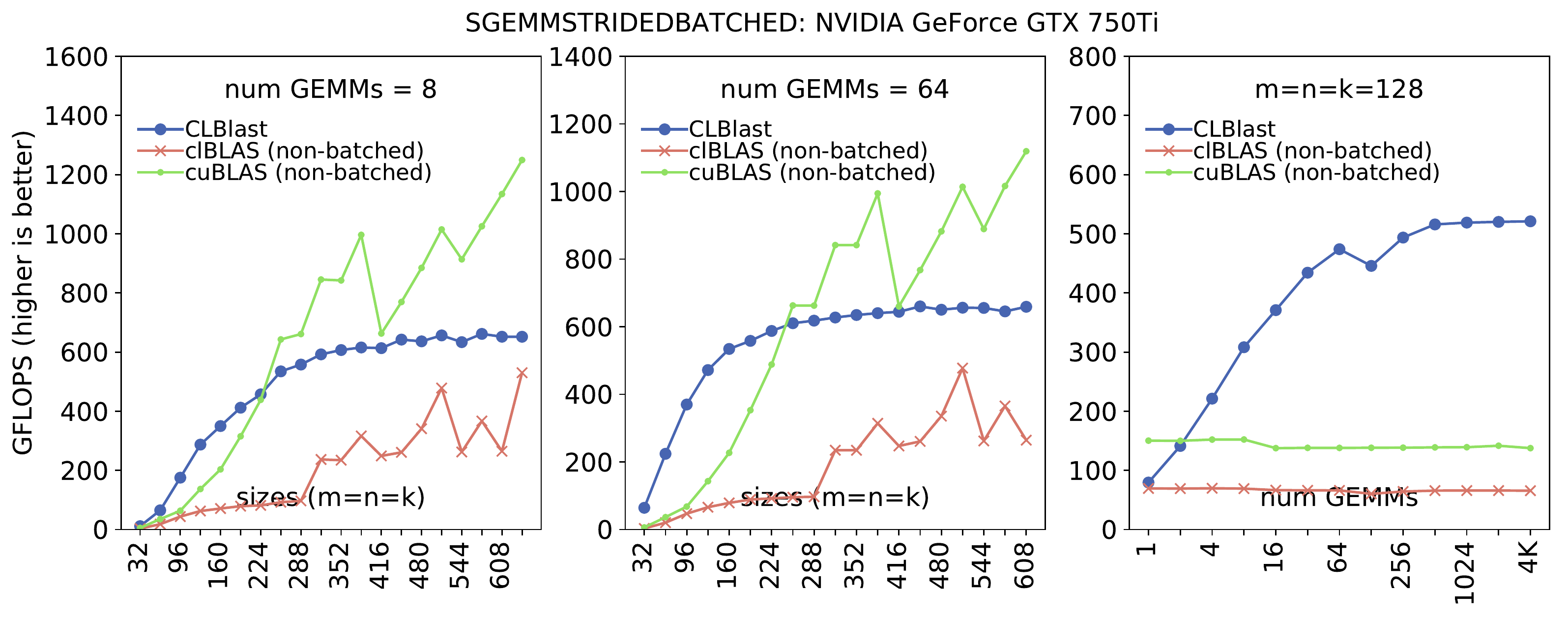}
  {\color{linecolor}\vline}
  \includegraphics[width=1.03\columnwidth]{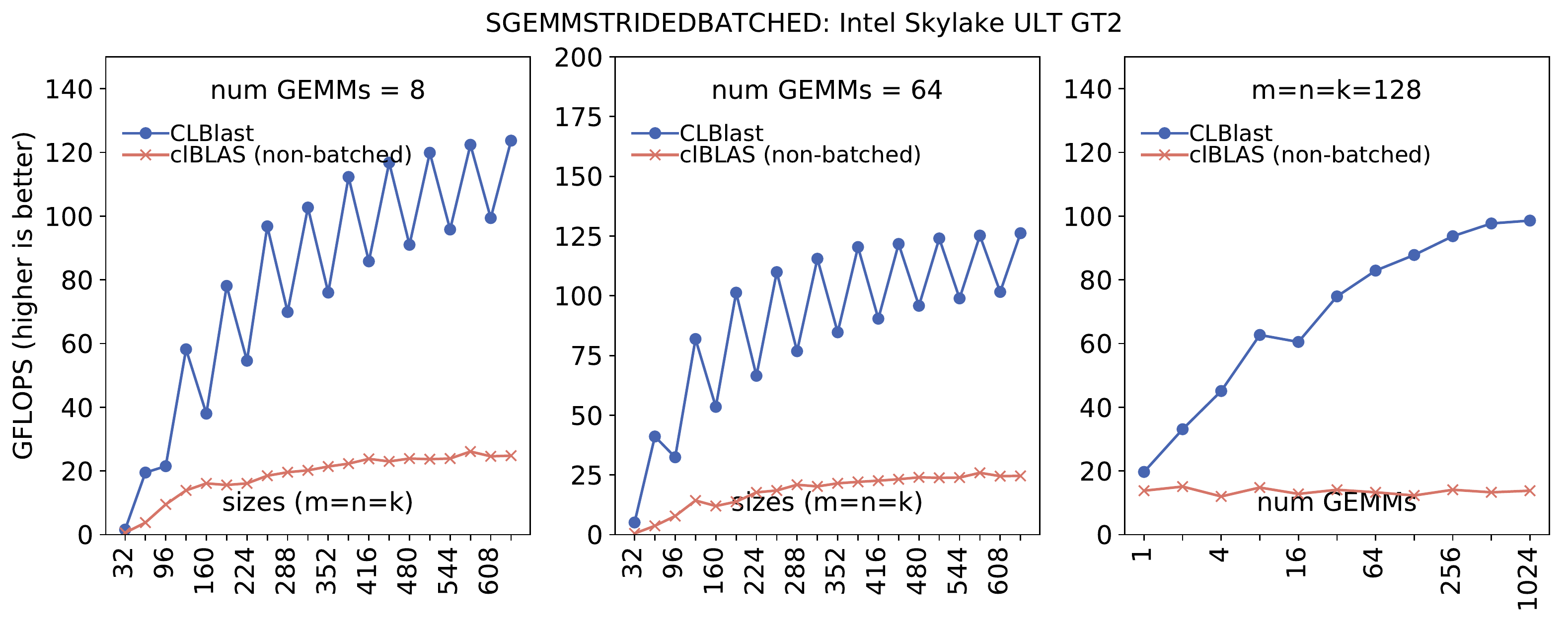}
  \caption{Performance of batched AXPY (top), generic batched GEMM (middle), and strided batched GEMM (bottom) on two different devices. Shown are results for the batched CLBlast routines (blue circles) and for running the non-batched routines from clBLAS (red crosses) and cuBLAS (green dots) in a loop. Note that cuBLAS also has a batched mode, but this is not included in this comparison.}
  \label{fig:batched}
\end{figure*}

Since CLBlast isn't originally designed for CUDA and since most CUDA devices can also run OpenCL applications, one might wonder about the advantages of a CUDA version of CLBlast? First of all, the library can now be integrated into existing CUDA projects, taking CUDA buffers directly as input. Secondly, it can now run on platforms for which NVIDIA does not ship an OpenCL implementation: all non-x86 systems such as the Jetson and Drive PX series and IBM Power based supercomputers. Finally, performance can be different from the OpenCL version. A first test showed that out-of-the-box the CUDA version was actually around 3\% slower for the SGEMM benchmarks on the GeForce GTX 750Ti test system. This could be due to different optimisation settings in NVIDIA's OpenCL/CUDA compiler. Nevertheless, the CUDA kernels also offer new optimisation opportunities such as using \texttt{\_\_ldg} or \texttt{\_\_shfl} intrinsics or mixed-precision tensor operations, all of which are not (yet) available in OpenCL. In this work we did not perform extensive CUDA versus OpenCL tests, but we leave this for future work.

\subsection{Batched BLAS Routines}

This section discusses the benefits of batched routines: grouping multiple similar traditional BLAS calls into a single routine for better efficiency. Although the overhead of the regular CLBlast routines is minimal, performing multiple small operations comes at a cost: the OpenCL device can become underutilized when running too few threads and work-groups. For example on an NVIDIA GTX 750 Ti with 5 compute units, a 32 by 32 matrix-multiplication with a work-group size of 512 results in 2 work-groups, leaving 3 units unoccupied. Even if the work-group size would be smaller, there would be insufficient threads to hide the GPU's memory latency. Batched BLAS routines can alleviate this issue by running unrelated but similar computations simultaneously.

CLBlast implements 3 batched routines: generic batched AXPY, generic batched GEMM, and strided batched GEMM. For the first two, their interfaces are similar to the non-batched counterpart, with the following exceptions:
\begin{enumerate}
  \item An additional parameter specifies the size of a batch.
  \item All offset arguments have become arrays, specifying the starting points of the individual vectors or matrices with respect to an OpenCL memory object for each of the individual computations within the batch.
  \item The scalar arguments (alpha and beta) are now arrays such that they can be set differently within the batch.
\end{enumerate}
The special strided batched GEMM is less generic: it doesn't have an offset array as argument. Instead, it has a single stride parameter for each of the matrices, specifying where the next data is. Also, it assumes alpha and beta to be equal across the whole batch. This variant is also implemented in cuBLAS and was shown by NVIDIA to reduce overhead significantly for small matrix sizes~\footnote{Blog on batching: \url{http://devblogs.nvidia.com/cublas-strided-batched-matrix-multiply}}.

These batched routines are specifically beneficial for machine learning applications, in which data is often processed in batches for training (minibatch stochastic gradient descent) and for inference (bulk-processing). In fact, deep-learning libraries such as cuBLAS/cuDNN and GreenTea libDNN~\cite{Tschopp2015} implement batched GEMM routines as well.

Figure~\ref{fig:batched} presents results of running batched AXPY and batched GEMM on two different devices. Additional results for batched routines can be found in the on-line appendix$^9$. We evaluate the benefit of batched routines compared to a non-batched clBLAS/cuBLAS reference for a fixed batch-size but with varying data-size (left and middle) and for a fixed data-size but with a varying batch-size (right). From the results, we conclude that the batched routines perform significantly better for small problems compared to their non-batched counter-parts (see also non-batched CLBlast in figure~\ref{fig:summary}). In some cases this can even be an order of magnitude better: batched routines can bring the performance of operations on small vectors and matrices to the level of much larger operations. Of course, the advantage of batching diminishes as input sizes grow bigger. We also note that the overhead due to loading the offsets and scalars is visible when comparing the generic batched GEMM with the strided version, which can be up to a factor 2 for the GTX 750Ti GPU, and even higher for the Skylake GPU. In the latter case we should note that these routines use the tuning parameters of the regular GEMM, which can be sub-optimal as shown for generic batched GEMM for small sizes. Thus, for optimal performance, tuning needs to be performed separately on the batched versions of the kernels.

\section{Future Work}

The current version of CLBlast is already production-ready and performs well on a variety of platforms. Nevertheless, we do identify topics of future work to further improve the library and the tuning:
\begin{itemize}
  \item The current out-of-the-box tuning parameters are optimized for specific routine arguments (e.g. matrix size). To get optimal performance for a particular use-case, the user is currently required to run the auto-tuner. However, if the tuners would be run per-default for a mixed set of arguments (e.g. both small and large matrices), we could estimate tuning parameters for every use-case. Selecting the argument mix and performing the estimation is not trivial and might require elaborate models of the kernels and the hardware, perhaps using machine learning such as in~\cite{Ballester2017,Falch2015}.
  \item The above can be applied in another dimension: predicting tuning parameters for unseen devices instead of for unseen arguments. Again, this might require sophisticated models of the kernels and hardware.
  \item The library already supports features useful for deep learning such as half-precision, batched GEMM, and an im2col implementation. However, further modifications can be made for deep learning, such as tensor-based convolutions and other cuDNN routines. Such additions would bring CLBlast in the scope of other auto-tuning work, such as~\cite{Tschopp2015,Moskewicz2017}.
\end{itemize}

\section{Conclusions}

This paper discussed CLBlast, an OpenCL BLAS library written in C++11. Thanks to its integrated auto-tuning support and the generic OpenCL kernels it performs well on a wide range of OpenCL devices including mobile and low-power GPUs, offering a viable alternative to the de-facto standard clBLAS or the closed-source cuBLAS. Users of the library can further improve performance by fine-tuning for their specific hardware or even for their specific use-case (e.g. matrix size). Integration into CUDA applications is also possible since the library also has a CUDA interface and back-end. Furthermore, this paper demonstrated that CLBlast is equipped with features which go beyond the standard BLAS definition: half-precision floating-point (FP16) and batched routines. Both are highly beneficial for deep-learning where batched operations are common and high precision is not always required. With the right hardware, CLBlast's FP16 mode can give you a factor two performance gain and memory savings. Batching can improve performance up to an order of magnitude depending on the use-case and hardware.

In conclusion, CLBlast is a production-ready and high-performance library which can be used today to accelerate code on OpenCL hardware. In the future CLBlast will continue to support the needs of the high-performance computing and deep learning communities to make the library even more tuned to the needs of the users. Readers are invited to contribute by sharing ideas for future work, tuning results or patches on \url{http://github.com/CNugteren/CLBlast}, the main project website of CLBlast.

\section*{Acknowledgments}

We thank all the contributors on the CLBlast project. We also thank \textit{dividiti} for providing access to ARM Mali hardware, \textit{Mark Wijtvliet} for access to an AMD GPU, and \textit{ArrayFire} for making continuous integration of CLBlast possible.


\bibliographystyle{ACM-Reference-Format}
\bibliography{references}

\end{document}